\documentclass[hyperref,showpacs,showkeys,12pt,titlepage]{revtex4}%
\usepackage{amsfonts}
\usepackage[debug,ps2pdf]{hyperref}
\usepackage{amsmath}
\usepackage{amssymb}
\usepackage[truetex]{graphicx}
\usepackage{bibmods}%
\providecommand{\U}[1]{\protect\rule{.1in}{.1in}}

\begin{document}
\title{Effect of the electric field on the creation of fermions in de-Sitter space-time}
\author{\textbf{S. Haouat}}
\email{s.haouat@gmail.com}
\affiliation{\textit{LPTh, Department of Physics, University of Jijel, BP 98, Ouled Aissa,
Jijel 18000, Algeria.}}
\author{\textbf{R. Chekireb}}
\affiliation{\textit{LPTh, Department of Physics, University of Jijel, BP 98, Ouled Aissa,
Jijel 18000, Algeria.}}

\begin{abstract}
The effect of the electric field on the creation of spin 1/2 particles from
vacuum in the (1+1) dimensional de-Sitter space-time is studied. The Dirac
equation with a constant electric field is solved by introducing an unitary
transformation. Then the canonical method based on Bogoliubov transformation
is applied to calculate the pair creation probability and the density of
created particles both for positive or negative wave vector. By doing
summation over all allowed states, the number of created particles per unit of
time per unit of length and the imaginary part of the Schwinger effective
action are expressed in closed forms. It is shown that the electric field
leads to a significant enhancement of the particle creation. The weak
expansion case and the limit $H=0$, where dS space reduces to the flat
Minkowski space-time, are discussed.

\end{abstract}

\maketitle

\newpage

\section{Introduction}

As we know our universe is undergoing an accelerated expansion that can be
approximated by an exponentially expanding de-Sitter space-time. It is widely
believed, also, that the de-Sitter space-time allows a better description of
the early stage of inflation. Therefore de-Sitter metric is of interest both
for the cosmology of the early and late universe. This importance comes also
from the fact that the de-Sitter space is the unique maximally symmetric
curved space and enjoys the same degrees of symmetry as Minkowski space-time.
This explains the increased interest to study physical quantum effects in such
a space-time. Among these effects we cite the particle-antiparticle pair
creation caused by the expansion of the universe \cite{FT1,FT2,FT3,FT4}. The
phenomenon of particle creation in de-Sitter space has been widely discussed
and analyzed from various point of view
\cite{ds1,ds2,ds3,ds4,ds5,ds6,ds7,ds8,ds9,ds10,ds11,ds12,ds13}. This is
because the fact that the particle creation has many important applications in
contemporary cosmology \cite{C1,C2,C3,C4,C5,C6,C7}. It is known that the
creation of particles induces an effective negative pressure which leads to an
inflationary phase with an exponential expansion. After a sufficient time of
inflation the exponential expansion will be adiabatically switched off giving
rise to a power law expansion. In addition, the creation of cold dark matter
by gravitational fields could explain the actual accelerated expansion without
appealing to the existence of dark energy. The particle creation mechanism
could also have significant contribution to the transition from an anisotropic
universe to an isotropic one \cite{VG}.

In de-Sitter space, the pair production rate, with a pure gravitational field,
is proportional to $\exp\left(  -\frac{2\pi}{H}m\right)  $, where $m$ is the
particle mass and $H$ is the Hubble's constant. This makes the production of
massive (heavy) particles negligible. The creation of massive particles is
then excluded to play any role in the evolution of the early universe. The
purpose of this paper is to study the effect of the electric field on the
creation of fermions from vacuum in the de-Sitter space-time. We know that
actually there is no electric field in the universe. However, as is mentioned
in \cite{Odintsov}, we reconcile that electric fields were present during the
initial stages of the formation of the universe and they vanish because of the
inverse effect of the particle creation. This idea resembles to the assumption
of the anisotropy of the early universe although the present universe is
isotropic. We note that the influence of electric field on the particle
creation in expanding universe has been studied in several models describing
different stages of the cosmic evolution
\cite{ds7,ds11,VG,Odintsov,Villalba1,haouat2,haouat3,PC1,PC2}.

The Schwinger effect in the (1+1) dimensional de-Sitter space-time is
extensively studied during the last few years \cite{2a,2b,2c,2d,2e,2f,2g}. The
geometric origin of the Schwinger effect in de-Sitter space is studied in
\cite{2c}. In \cite{2d}, the authors studied the Schwinger mechanism by a
uniform electric field in $dS_{2}$ and $AdS_{2}$. They expressed the one-loop
effective action in the proper-time integral. Then they computed the imaginary
part of the effective action and the pair production rate. The authors of
\cite{2e}, considered the problem of particle production by a constant
electric field in (1+1) dimensional de Sitter space as a model for describing
false vacuum decay beyond the semiclassical approximation. They found that the
adiabatic \textquotedblleft in\textquotedblright\ vacuum associated with the
flat chart develops a space-like expectation value for the current, which
manifestly breaks the de Sitter invariance of the background fields. However
the case studied in particular detail is that of scalar particles. Less
studied is the case of fermions, where the spin of the fermion could play an
important role on the behavior of created particles. An interesting attempt to
study the creation of fermions by an electric field in de-Sitter space is
communicated several years ago in \cite{2f} (see also \cite{2g}).

In order to study the effect of particle-antiparticle pair creation form
vacuum by gravitational and electric fields we have at our disposal several
methods such as the adiabatic method \cite{Parker1,Parker2}, the Hamiltonian
diagonalization technique \cite{Grib1,Grib2}, the Feynman path integral
derivation \cite{Chitre,Duru}, the Green function approach
\cite{Bukhbinder,Gavrilov}, the semiclassical WKB approximation
\cite{Biswas1,Biswas2,Biswas3} as well as the method based on Bogoliubov
transformation \cite{Villalba2,haouat1} that we shall use in this paper. The
later method is more convenient for the spin $\frac{1}{2}$ case.

The paper is organized as follows; At the beginning we consider a spin
$\frac{1}{2}$\ fermion subjected to an electric field in the (1 + 1)
dimensional de-Sitter space-time. Then we solve the corresponding Dirac
equation by introducing an unitary transformation. To investigate the process
of particle creation we apply the canonical method based on Bogoliubov
transformation connecting the "\textit{in}" with the "\textit{out}" states.
This method permits us to determine the probability to create pair of
particles in a given state, the density number of created particles and the
vacuum persistence. To clarify the effect of the electric field on the
creation of particles we compute the number of created particles per unit of
time per unit of length by doing sum over all states and we show how the
electric field influences on the particle creation. We complete the analysis
by deriving the imaginary part of the Schwinger effective Lagrangian starting
from the vacuum to vacuum transition amplitude. These results allows us also
to discuss the effect of the actual expansion of the universe on the Schwinger
effect. If the universe expansion leads to some enhancement to the Schwinger
effect, this would be of great interest to experimental physics. We shall
discuss this issue.

\section{Dirac equation in an expanding universe}

Let us consider a spin $\frac{1}{2}$\ fermion of mass $m$ and charge $\left(
-e\right)  $ moving on the background geometry of a (1 + 1) dimensional
expanding universe described by the line element%
\begin{equation}
ds^{2}=a^{2}\left(  \eta\right)  \left(  d\eta^{2}-dx^{2}\right)  , \label{2}%
\end{equation}
where $a\left(  \eta\right)  =a_{0}f\left(  \eta\right)  $, in the presence of
an electric field described by the vector potential $A_{\mu}=(0,A_{1}\left(
\eta\right)  )$, with
\begin{equation}
A_{1}\left(  \eta\right)  =E_{0}f\left(  \eta\right)  . \label{3}%
\end{equation}
Here $f\left(  \eta\right)  $ is an arbitrary function. The covariant Dirac
equation with the metric (\ref{2}) takes the form%
\begin{equation}
\left[  i\tilde{\gamma}^{\mu}\left(  \eta\right)  (\partial_{\mu}-ieA_{\mu
}(\eta)-\Gamma_{\mu}(\eta))-m\right]  \psi=0 \label{4}%
\end{equation}
where the curvature-dependent Dirac matrices $\tilde{\gamma}^{\mu}(\eta)$ are
given in diagonal tetrad gauge by
\begin{align}
\tilde{\gamma}^{0}(\eta)  &  =\frac{1}{a\left(  \eta\right)  }\gamma
^{0}\label{5}\\
\tilde{\gamma}^{1}(\eta)  &  =-\frac{1}{a\left(  \eta\right)  }\gamma^{1}
\label{6}%
\end{align}
where $\gamma^{0}$ and $\gamma^{1}$ are the usual Dirac's matrices that can be
written, in (1+1) dimensional Minkowski space, in terms of Pauli matrices as
follows%
\begin{equation}
\gamma^{0}=\sigma_{z}~,~\gamma^{1}=i\sigma_{y} \label{7}%
\end{equation}
and $\Gamma_{\mu}(\eta)$ are the spin connections%
\begin{align}
\Gamma_{0}  &  =0\nonumber\\
\Gamma_{1}  &  =\frac{1}{2}\frac{\dot{a}(\eta)}{a\left(  \eta\right)  }%
\gamma_{0}\gamma_{1}. \label{8}%
\end{align}
By making the substitution $\chi\left(  \eta,x\right)  =a^{\frac{1}{2}}%
(\eta)\psi\left(  \eta,x\right)  ,$ we obtain the simpler equation%
\begin{equation}
\left[  \gamma^{\mu}\left(  i\partial_{\mu}+eA_{\mu}(\eta)\right)  -ma\left(
\eta\right)  \right]  \chi\left(  \eta,x\right)  =0. \label{9}%
\end{equation}
In order to solve this equation we write, at the beginning, $\chi\left(
\eta,x\right)  =\exp\left(  ikx\right)  \xi\left(  \eta\right)  $ with%
\begin{equation}
\xi\left(  \eta\right)  =\left(
\begin{array}
[c]{c}%
\xi_{1}\left(  \eta\right) \\
\xi_{2}\left(  \eta\right)
\end{array}
\right)  \label{10}%
\end{equation}
where the two components $\xi_{1}\left(  \eta\right)  $ and $\xi_{2}\left(
\eta\right)  $ satisfy the two coupled equations
\begin{align}
\left(  i\frac{\partial}{\partial\eta}-ma_{0}f\left(  \eta\right)  \right)
\xi_{1}\left(  \eta\right)   &  =\left[  k-eE_{0}f\left(  \eta\right)
\right]  \xi_{2}\left(  \eta\right) \\
\left(  i\frac{\partial}{\partial\eta}+ma_{0}f\left(  \eta\right)  \right)
\xi_{2}\left(  \eta\right)   &  =\left[  k-eE_{0}f\left(  \eta\right)
\right]  \xi_{1}\left(  \eta\right)  .
\end{align}
Here we remark that the coupling coefficient $\left[  k-eE_{0}f\left(
\eta\right)  \right]  $ depends on the conformal time and the usual iteration
procedure leads to a complicated second order equation that does not admit
well-known solutions for many models. To simplify the problem let us introduce
the unitary transformation
\begin{equation}
\left(
\begin{array}
[c]{c}%
\xi_{1}\left(  \eta\right) \\
\xi_{2}\left(  \eta\right)
\end{array}
\right)  =\frac{1}{\sqrt{1+\tau^{2}}}\left(
\begin{array}
[c]{cc}%
1 & \tau\\
-\tau & 1
\end{array}
\right)  \left(
\begin{array}
[c]{c}%
\varphi_{1}\left(  \eta\right) \\
\varphi_{2}\left(  \eta\right)
\end{array}
\right)  \label{13}%
\end{equation}
with
\begin{equation}
\tau=\frac{a_{0}}{eE_{0}}\left(  \mathcal{M}-m\right)  \label{14}%
\end{equation}
and%
\begin{equation}
\mathcal{M}=\sqrt{m^{2}+\frac{e^{2}E_{0}^{2}}{a_{0}^{2}}}. \label{15}%
\end{equation}
Then the novel components $\varphi_{1}\left(  \eta\right)  $ and $\varphi
_{2}\left(  \eta\right)  $ satisfy the following system of equations
\begin{align}
\left[  i\frac{\partial}{\partial\eta}-\mathcal{M}a_{0}f\left(  \eta\right)
+\frac{eE_{0}}{\mathcal{M}a_{0}}k\right]  \varphi_{1}\left(  \eta\right)   &
=k\frac{m}{\mathcal{M}}\varphi_{2}\left(  \eta\right) \label{16a}\\
\left[  i\frac{\partial}{\partial\eta}+\mathcal{M}a_{0}f\left(  \eta\right)
-\frac{eE_{0}}{\mathcal{M}a_{0}}k\right]  \varphi_{2}\left(  \eta\right)   &
=k\frac{m}{\mathcal{M}}\varphi_{1}\left(  \eta\right)  \label{17a}%
\end{align}
which leads to the second order equation%
\begin{equation}
\left[  \frac{\partial^{2}}{\partial\eta^{2}}+\mathcal{M}^{2}a_{0}^{2}%
f^{2}\left(  \eta\right)  \pm i\mathcal{M}a_{0}f^{\prime}\left(  \eta\right)
-2keE_{0}f\left(  \eta\right)  +k^{2}\right]  \varphi_{1,2}\left(
\eta\right)  =0. \label{se}%
\end{equation}
Equation (\ref{se}) admits simple and analytic solution for various models,
such as the de-Sitter space with $f\left(  \eta\right)  =\frac{-1}{H^{2}\eta}$
the radiation dominated universe for $f\left(  \eta\right)  =\eta$ and the
Milne universe for $f\left(  \eta\right)  =e^{\rho\eta}$, ...etc.

Let us notice that the gravitational field couples to the mass of the
particle, while the electric field couples to the charge. In the novel system
we can see that an effective field is coupled to the quantity $\mathcal{M}$.
As we will show, for the process of particle creation, this quantity is more
important than the mass of the particle.

Writing the equation (\ref{se}) in the form $\varphi_{s}^{\prime\prime}\left(
\eta\right)  +\omega_{s}^{2}(\eta)\varphi_{s}\left(  \eta\right)  =0$, with
\begin{equation}
\omega_{s}^{2}(\eta)=\mathcal{M}^{2}a_{0}^{2}f^{2}\left(  \eta\right)  \pm
i\mathcal{M}a_{0}f^{\prime}\left(  \eta\right)  -2keE_{0}f\left(  \eta\right)
+k^{2},
\end{equation}
we find that the particles creation is well-defined only in the adiabatic
condition%
\begin{equation}
\lim\limits_{\eta\rightarrow\eta_{0}}\left\vert \frac{\dot{\omega}_{s}(\eta
)}{\omega_{s}^{2}(\eta)}\right\vert <<1 \label{19}%
\end{equation}
In addition, at the limit $m\rightarrow0$, the mixing term in (\ref{16a}) and
(\ref{17a}) vanishes and the positive and negative energy solutions never
intercept each other. This means that there is no production of massless
particles even if an electric field is present. In such a case we should treat
matter as waves rather than particles.

Furthermore, equation (\ref{se}) can be written in the form%
\begin{equation}
\left[  \frac{\partial^{2}}{\partial\eta^{2}}+\left[  \tilde{k}-e\mathcal{E}%
f\left(  \eta\right)  \right]  ^{2}+\tilde{M}^{2}\pm ie\mathcal{E}f^{\prime
}\left(  \eta\right)  \right]  \varphi_{1,2}\left(  \eta\right)  =0,
\label{efe}%
\end{equation}
where%
\begin{align}
e\mathcal{E}  &  =\mathcal{M}a_{0},\\
\tilde{M}^{2}  &  =k^{2}-\tilde{k}^{2}=k^{2}\frac{m^{2}}{\mathcal{M}^{2}}%
\end{align}
and%
\begin{equation}
\tilde{k}=k\frac{eE_{0}}{\mathcal{M}a_{0}}.
\end{equation}
Equation (\ref{efe}) is similar to the quadratic Dirac equation for a particle
of mass $\tilde{M}$, charge $\left(  -e\right)  $ and wave vector $\tilde{k}$
interacting with the gauge field $A_{1}=\mathcal{E}f\left(  \eta\right)  $ in
Minkowski space. Then the density of particles with mass $m$, charge $\left(
-e\right)  $ and wave vector $k$, created by an electric field $A_{1}%
=E_{0}f\left(  \eta\right)  $ in an expanding universe with the scale factor
$a_{0}f\left(  \eta\right)  $ is the same as the density of particles with
mass $\tilde{M}$, charge $\left(  -e\right)  $ and wave vector $\tilde{k}$
created by the electric field $A_{1}=\mathcal{E}f\left(  \eta\right)  $ in
Minkowski space.

Technically, the study of particle creation in an expanding universe requires
a definition of a vacuum state for the field theory \cite{Winitzki}. However,
unlike the free Dirac field, it is not obvious how to determine the vacuum
states when the spinor field is subjected to a general gravitational
background. It is widely believed that in arbitrary curved background, there
is no absolute definition of the vacuum state and the concept of particles is
not completely clear. From physical point of view, it is well known that in
the standard quantum theory a particle cannot be localized to a region smaller
than its de Broglie wavelength. When this wavelength is sufficiently large,
the concept of particle becomes unclear \cite{Anton}. Furthermore, when the
vacuum state is defined in the remote past it is habitually unstable so that
it may differ from the vacuum state in the remote future. This gives rise to
spontaneous particle creation. The derivation of the effect requires also
exact solutions to the field equation. If these solutions are classified as
"in" and "out" states then the Bogoliubov transformation between these states
leads to exact expressions for the number density of created particles and the
probability to create a pair of particles in a given state. In the next
section we consider the creation of particle-antiparticle pairs in a de-Sitter
space with a constant electric field.

\section{Pair creation in de-Sitter space}

The general technique for solving the Dirac equation in a (1+1) dimensional
expanding universe when an electric field is present being shown, let us
consider the important case of de-Sitter space with a constant electric field.
The line element of the metric describing $dS_{2}$ space-time can be written
as%
\begin{equation}
ds^{2}=dt^{2}-e^{2Ht}dx^{2}=a^{2}\left(  \eta\right)  \left(  d\eta^{2}%
-dx^{2}\right)  ,
\end{equation}
where $a\left(  \eta\right)  =\frac{-1}{H\eta}$ and $H$ is the Hubble's
constant. A constant electric field in the comoving system of coordinates is
described by the vector potential
\begin{equation}
A_{1}\left(  \eta\right)  =-\frac{E_{0}}{H^{2}\eta}.
\end{equation}
This is just a particular case of the problem studied in the previous section,
with $a_{0}=H$, $\mathcal{M}=\sqrt{m^{2}+\frac{e^{2}E_{0}^{2}}{H^{2}}}$ and
$f\left(  \eta\right)  =-\frac{1}{H^{2}\eta}$. In such a case, the equation
(\ref{se}) is similar to the well-known Whittaker equation \cite{Grad}%
\begin{equation}
\left[  \frac{d^{2}}{d\rho^{2}}-\frac{1}{4}+\frac{\lambda}{\rho}+\frac
{\frac{1}{4}-\mu_{s}^{2}}{\rho^{2}}\right]  \tilde{\varphi}_{s}\left(
\rho\right)  =0, \label{21}%
\end{equation}
with%
\begin{equation}
\rho=-2ik\eta\label{20}%
\end{equation}
and $\tilde{\varphi}_{s}\left(  \rho\right)  \equiv\varphi_{s}\left(
\eta\right)  $, $s=\overline{1,2}$. The constants $\mu_{s}$ and $\lambda$ are
given by
\begin{align}
\mu_{1}  &  =\frac{1}{2}-i\frac{\mathcal{M}}{H}=\mu\nonumber\\
\mu_{2}  &  =\frac{1}{2}+i\frac{\mathcal{M}}{H}=\mu^{\ast}=1-\mu\label{22}%
\end{align}
and%
\begin{equation}
\lambda=i\frac{eE_{0}}{H^{2}}. \label{23}%
\end{equation}
It is known that one can find for the equation (\ref{21}) several sets of
linearly independent solutions which can be written in terms of the Whittaker
functions $M_{\lambda,\mu}\left(  \rho\right)  $ and $W_{\lambda,\mu}\left(
\rho\right)  $, with%
\begin{align}
M_{\lambda,\mu}\left(  \rho\right)   &  =\rho^{\mu+\frac{1}{2}}e^{-\frac{\rho
}{2}}M\left(  \mu-\lambda+\frac{1}{2},2\mu+1;\rho\right) \label{24}\\
W_{\lambda,\mu}\left(  \rho\right)   &  =\rho^{\mu+\frac{1}{2}}e^{-\frac{\rho
}{2}}U\left(  \mu-\lambda+\frac{1}{2},2\mu+1;\rho\right)  , \label{25}%
\end{align}
where $M\left(  a,b,\rho\right)  $ and $U\left(  a,b,\rho\right)  $ are the
Kummar functions \cite{Abramo}.

To obtain a well-defined vacuum state with a reasonable choice of positive and
negative frequency modes we use the so-called adiabatic method based on the
solutions of the relativistic Hamilton-Jacobi equation. Taking into account
the asymptotic behavior of the $W_{\lambda,\mu}\left(  \rho\right)  $ functions%

\begin{equation}
W_{\lambda,\mu}\left(  \rho\right)  \sim e^{-\frac{\rho}{2}}\left(
-\rho\right)  ^{\lambda} \label{26}%
\end{equation}
and using the solutions of the Hamilton-Jacobi equation we can find for the
"\textit{in}" states the following expressions%
\begin{equation}
\xi_{in}^{+}\left(  \eta\right)  =\mathcal{N}\left(
\begin{array}
[c]{c}%
\sqrt{\mathcal{M}-\frac{eE_{0}}{H}}W_{-\lambda,\mu}\left(  -\rho\right)
+\tau\sqrt{\mathcal{M}+\frac{eE_{0}}{H}}W_{-\lambda,1-\mu}\left(  -\rho\right)
\\
-\tau\sqrt{\mathcal{M}-\frac{eE_{0}}{H}}W_{-\lambda,\mu}\left(  -\rho\right)
+\sqrt{\mathcal{M}+\frac{eE_{0}}{H}}W_{-\lambda,1-\mu}\left(  -\rho\right)
\end{array}
\right)  \label{32}%
\end{equation}
and%
\begin{equation}
\xi_{in}^{-}\left(  \eta\right)  =\mathcal{N}^{\ast}\left(
\begin{array}
[c]{c}%
\sqrt{\mathcal{M}+\frac{eE_{0}}{H}}W_{\lambda,\mu}\left(  \rho\right)
-\tau\sqrt{\mathcal{M}-\frac{eE_{0}}{H}}W_{\lambda,1-\mu}\left(  \rho\right)
\\
\tau\sqrt{\mathcal{M}+\frac{eE_{0}}{H}}W_{\lambda,\mu}\left(  \rho\right)
+\sqrt{\mathcal{M}-\frac{eE_{0}}{H}}W_{\lambda,1-\mu}\left(  \rho\right)
\end{array}
\right)  \label{33}%
\end{equation}
$\allowbreak$where $\mathcal{N}$ is a normalization constant that is
unimportant vis-a-vis the mechanism of particle creation.

Let us now define the "\textit{out}" states. These states can be defined by
studying the limit $\eta\rightarrow0$ $(t\rightarrow+\infty)$. However, it
should be noted that, since the de-Sitter space describes the universe in the
inflationary era in a limited period, the study of "out" states by considering
the limit $\eta\rightarrow0$ is physically inappropriate. This note is missing
in several derivations of this effect in the literature, where the probability
to create a pair of particles and the number density of created particles were
obtained by studying the limit $\eta\rightarrow0$.

Taking into account that the Kummer function $M\left(  a,b;\rho\right)  $ is
by definition%

\begin{equation}
M\left(  a,b;\rho\right)  =1+\frac{a}{1}\frac{\rho}{b}+\frac{a\left(
a+1\right)  }{1\times2}\frac{\rho^{2}}{b\left(  b+1\right)  }+\frac{a\left(
a+1\right)  \left(  a+2\right)  }{1\times2\times3}\frac{\rho^{3}}{b\left(
b+1\right)  \left(  b+2\right)  }+...
\end{equation}
we find that, for finite $\rho$, $M\left(  a,b;\rho\right)  \approx1$ if
$\frac{\left\vert \rho\right\vert }{\left\vert b\right\vert }<<1$. Then,
instead of taking the limit $\eta\rightarrow0$ $(t\rightarrow+\infty)$, we
consider the case when
\begin{equation}
\left\vert \rho\right\vert <<\left\vert 2\mu+1\right\vert .
\end{equation}
In such a case, like at $\rho\rightarrow0$, the functions $M_{\lambda,\mu
}\left(  \rho\right)  $\ have the asymptotic behavior
\begin{equation}
M_{\lambda,\mu}\left(  \rho\right)  \sim e^{-\frac{\rho}{2}}\rho^{\mu+\frac
{1}{2}}. \label{36}%
\end{equation}
Consequently, we arrive at the following expressions for the two components of
the Dirac spinor
\begin{equation}
\xi_{out}^{+}\left(  \eta\right)  =\mathcal{N}^{\prime}\left(
\begin{array}
[c]{c}%
M_{\lambda,-\mu}\left(  \rho\right)  +\tau\frac{\frac{m}{\mathcal{M}}%
}{4\left(  \frac{1}{2}+i\frac{\mathcal{M}}{H}\right)  }M_{\lambda,-\mu
+1}\left(  \rho\right) \\
-\tau M_{\lambda,-\mu}\left(  \rho\right)  +\frac{\frac{m}{\mathcal{M}}%
}{4\left(  \frac{1}{2}+i\frac{\mathcal{M}}{H}\right)  }M_{\lambda,-\mu
+1}\left(  \rho\right)
\end{array}
\right)  \label{41}%
\end{equation}
and%
\begin{equation}
\xi_{out}^{-}\left(  \eta\right)  =\mathcal{N}^{\prime\ast}\left(
\begin{array}
[c]{c}%
-\tau M_{-\lambda,\mu-1}\left(  -\rho\right)  +\frac{\frac{m}{\mathcal{M}}%
}{4\left(  \frac{1}{2}-i\frac{\mathcal{M}}{H}\right)  }M_{-\lambda,\mu}\left(
-\rho\right) \\
M_{-\lambda,\mu-1}\left(  -\rho\right)  +\tau\frac{\frac{m}{\mathcal{M}}%
}{4\left(  \frac{1}{2}-i\frac{\mathcal{M}}{H}\right)  }M_{-\lambda,\mu}\left(
-\rho\right)
\end{array}
\right)  . \label{43}%
\end{equation}
Note that those states are connected to one another by the charge conjugation
transformation defined by
\begin{equation}
\xi\rightarrow\xi^{c}=\sigma_{1}\xi^{\ast}. \label{34}%
\end{equation}
In addition positive and negative energy solutions satisfy the orthogonality
condition%
\begin{equation}
\bar{\xi}^{+}\xi^{-}=\bar{\xi}^{-}\xi^{+}=0. \label{35}%
\end{equation}

Now, as we have mentioned above, in order to determine the probability of pair
creation and the density of created particles we use the Bogoliubov
transformation connecting the "\textit{in}" with the "\textit{out}" states.
The relation between those states can be obtained by the use of the relation
between Whittaker functions \cite{Grad}
\begin{equation}
M_{\lambda,\mu}\left(  z\right)  =\frac{\Gamma\left(  2\mu+1\right)
e^{i\pi\lambda}}{\Gamma\left(  \mu-\lambda+\frac{1}{2}\right)  }%
W_{-\lambda,\mu}\left(  -z\right)  +\frac{\Gamma\left(  2\mu+1\right)
e^{-i\pi\left(  \mu-\lambda+\frac{1}{2}\right)  }}{\Gamma\left(  \mu
+\lambda+\frac{1}{2}\right)  }W_{\lambda,\mu}\left(  z\right)  , \label{45}%
\end{equation}
with $-\frac{3\pi}{2}<\arg z<\frac{\pi}{2}$ and $2\mu\neq-1,-2,\cdot\cdot
\cdot$. The functions $\xi_{out}^{+}\left(  \eta\right)  $ and $\xi_{out}%
^{+}\left(  \eta\right)  $ can be then expressed in terms of $\xi_{in}%
^{+}\left(  \eta\right)  $ and $\xi_{in}^{-}\left(  \eta\right)  $ as follows%
\begin{equation}%
\begin{array}
[c]{c}%
\xi_{out}^{+}\left(  \eta\right)  =\alpha~\xi_{in}^{+}\left(  \eta\right)
+\beta~\xi_{in}^{-}\left(  \eta\right) \\
\xi_{out}^{-}\left(  \eta\right)  =\alpha^{\ast}~\xi_{in}^{-}\left(
\eta\right)  +\beta^{\ast}~\xi_{in}^{+}\left(  \eta\right)
\end{array}
\label{46}%
\end{equation}
where the Bogoliubov coefficients $\alpha~$and $\beta$ are given by
\begin{equation}
\frac{\beta}{\alpha}=\frac{\mathcal{N}}{\mathcal{N}^{\ast}}\frac{\Gamma\left(
\frac{1}{2}-\mu-\lambda\right)  \sqrt{\mathcal{M}-\frac{eE_{0}}{H}}}%
{\Gamma\left(  \frac{1}{2}-\mu+\lambda\right)  \sqrt{\mathcal{M}+\frac{eE_{0}%
}{H}}}e^{-i\pi\left(  \mu-\frac{1}{2}\right)  } \label{47}%
\end{equation}
and%
\begin{equation}
\left\vert \alpha\right\vert ^{2}+\left\vert \beta\right\vert ^{2}=1.
\label{48}%
\end{equation}
The Bogoliubov relation between "\textit{in}" and "\textit{out}" states leads
to a relation between the creation and annihilation operators
\begin{equation}%
\begin{array}
[c]{c}%
\hat{b}_{-k,in}^{+}=\alpha^{\ast}~\hat{b}_{-k,out}^{+}+\beta~\hat{a}_{k,out}\\
\hat{a}_{k,in}=\alpha~\hat{a}_{k,out}+\beta^{\ast}~\hat{b}_{-k,out}^{+}.
\end{array}
\label{49}%
\end{equation}
Therefore, the probability of pair creation and the density of created
particles will be given in terms of Bogoliubov coefficients. In effect
starting from the amplitude $\mathcal{A}=\left\langle 0_{out}\left\vert
a_{k,out}b_{-k,out}\right\vert 0_{in}\right\rangle $, it is easy to show that%
\begin{equation}
\mathcal{A}=-\frac{\beta^{\ast}}{\alpha}\left\langle 0_{out}\right\vert
\left.  0_{in}\right\rangle . \label{50}%
\end{equation}
The probability to create a pair of fermions in the state $k$ is then%
\begin{equation}
\mathcal{P}_{k}=\left\vert \frac{\beta}{\alpha}\right\vert ^{2}. \label{51}%
\end{equation}
By the use of the following property of gamma function \cite{Grad}
\begin{equation}
\left\vert \Gamma(ix)\right\vert ^{2}=\frac{\pi}{x\sinh\pi x}%
\end{equation}
we find%
\begin{equation}
\left\vert \frac{\beta}{\alpha}\right\vert ^{2}=\frac{\sinh\pi\left(
\frac{\mathcal{M}}{H}+\frac{eE_{0}}{H^{2}}\right)  }{\sinh\pi\left(
\frac{\mathcal{M}}{H}-\frac{eE_{0}}{H^{2}}\right)  }e^{-2\pi\frac{\mathcal{M}%
}{H}}. \label{52}%
\end{equation}
For the number density of created particles we have%
\begin{equation}
n\left(  k\right)  =\left\langle 0_{in}\left\vert a_{k,out}^{+}a_{k,out}%
\right\vert 0_{in}\right\rangle =\left\vert \beta\right\vert ^{2}. \label{53}%
\end{equation}
Taking into account that the Bogoliubov coefficients satisfy the condition
(\ref{48}), it is easy to show that%
\begin{equation}
n\left(  k\right)  =\frac{\sinh\pi\left(  \frac{\mathcal{M}}{H}+\frac{eE_{0}%
}{H^{2}}\right)  e^{-2\pi\frac{\mathcal{M}}{H}}}{\sinh\pi\left(
\frac{\mathcal{M}}{H}-\frac{eE_{0}}{H^{2}}\right)  +\sinh\pi\left(
\frac{\mathcal{M}}{H}+\frac{eE_{0}}{H^{2}}\right)  e^{-2\pi\frac{\mathcal{M}%
}{H}}}. \label{54}%
\end{equation}
Let us note that these results are obtained for a positive wave vector ($k>0$)
and $-\frac{3\pi}{2}<\arg\left(  2ik\eta\right)  <\frac{\pi}{2}$. For the the
case when $k<0$, the quantities $n\left(  k\right)  $ and $\mathcal{P}_{k}$
can be obtained form (\ref{52}) and (\ref{54}) by changing the sign of $e$. We
have then%
\begin{equation}
n\left(  k\right)  =\frac{\sinh\pi\left(  \frac{\mathcal{M}}{H}%
+\operatorname{sign}\left(  k\right)  \frac{eE_{0}}{H^{2}}\right)
e^{-2\pi\frac{\mathcal{M}}{H}}}{\sinh\pi\left(  \frac{\mathcal{M}}%
{H}-\operatorname{sign}\left(  k\right)  \frac{eE_{0}}{H^{2}}\right)
+\sinh\pi\left(  \frac{\mathcal{M}}{H}+\operatorname{sign}\left(  k\right)
\frac{eE_{0}}{H^{2}}\right)  e^{-2\pi\frac{\mathcal{M}}{H}}}%
\end{equation}
$\allowbreak$ Then $n\left(  k\right)  $ is more significant when $k>0$.
Therefore the constant electric field produces predominantly particles with
$k>0$. In other wards, in the presence of a constant electric field, particles
prefer to be created in a specific direction. This depends on the orientation
of the electric field and the sign of the particle charge. The antiparticles
are in the main created in the opposite direction with the same density as the
particles. $\allowbreak$

Since the adiabatic condition (\ref{19}) reduces to $\frac{H}{\mathcal{M}}%
<<1$, the particle creation in de-Sitter space is well-defined only if
$\mathcal{M}>>H$. In such a limit, the density of created particles can be
approximated by
\begin{equation}
n\left(  k\right)  =\exp\left[  -\frac{2\pi}{H}\left(  \mathcal{M}%
-\operatorname{sign}\left(  k\right)  \frac{eE_{0}}{H}\right)  \right]  ,
\label{55}%
\end{equation}
which resembles to the Boltzmann distribution. This explains the thermal
nature of the effect and shows that the spin effects are negligible at this level.

\section{The number of created particles}

Since the electric field amplifies the particle creation only when $k>0$, it
is of interest to discuss the effect of the electric field on the total number
of created particles. The total number of created particles is given by
\begin{equation}
N=\int\frac{dxdk}{2\pi a\left(  t\right)  }n\left(  k\right)  \label{nt}%
\end{equation}
where $\frac{dxdk}{2\pi a\left(  t\right)  }$ is the number of states in the
phase space element $dxdk$ and the factor $a\left(  t\right)  $ in the
denominator expresses the dilution of the particles by the expansion of the
universe. Incorporating equation (\ref{55}) into (\ref{nt}) we get
\begin{equation}
N=\frac{1}{\pi a\left(  t\right)  }\int dx\int_{k\geq0}dk\exp\left(
-\frac{2\pi}{H}\mathcal{M}\right)  \cosh\left(  2\pi\frac{eE_{0}}{H^{2}%
}\right)  \label{n}%
\end{equation}
Since $n\left(  k\right)  $ depends only on $\operatorname{sign}\left(
k\right)  $, the integral over $k$ is divergent. The origin of this divergence
is the fact that the total number of created particles in an infinite time is
infinite. However, the number of created particles per unit of time per unit
of length $\frac{dN}{dxdt}$ must be finite. This quantity which is directly
related to the experimental measurements, is defined by
\begin{equation}
N=\int dN=\int\frac{dN}{dxdt}dxdt.
\end{equation}
Then, divergence arises in the integration $\int dxdk$ must be equivalent to
the divergence arise in the integration over $\int dxdt$. In the flat
Minkowski space-time the integration over $k$ can be carried out by making the
change $\int dk\rightarrow eE_{0}\int dt=eE_{0}T$. In de-Sitter space, the
situation is slightly different. By taking into account that particles and
antiparticle are well-defined only when $\left\vert \rho\right\vert
<<\left\vert 2\mu+1\right\vert $, we find that $n\left(  k\right)  $ makes
sense only if
\begin{equation}
\left\vert k\right\vert <\mathcal{M}e^{Ht}.
\end{equation}
Thus, at any given time $t$ we only need to integrate up to a cut-off
$\Lambda=\mathcal{M}e^{Ht}$. If we consider the variation of $N$ with respect
to a small variation of the cut-off, we obtain%
\begin{equation}
\frac{dN}{d\Lambda}=\frac{1}{\pi a\left(  t\right)  }\int dx\exp\left(
-\frac{2\pi}{H}\mathcal{M}\right)  \cosh\left(  2\pi\frac{eE_{0}}{H^{2}%
}\right)  .
\end{equation}
Since
\begin{equation}
d\Lambda=\mathcal{M}He^{Ht}dt,
\end{equation}
we obtain%
\begin{equation}
\frac{dN}{dt}=\frac{\mathcal{M}H}{\pi}\int dx\cosh\left(  2\pi\frac{eE_{0}%
}{H^{2}}\right)  \exp\left(  -\frac{2\pi}{H}\mathcal{M}\right)
\end{equation}
and, consequently, the number of created particles per unit of time per unit
of length is given by%
\begin{equation}
\frac{dN}{dxdt}=\frac{\mathcal{M}H}{\pi}\cosh\left(  2\pi\frac{eE_{0}}{H^{2}%
}\right)  \exp\left(  -\frac{2\pi}{H}\mathcal{M}\right)  . \label{67}%
\end{equation}
Here, we notice that, in de-Sitter space-time, the integration over $k$ can be
simply replaced by
\begin{equation}
\int\frac{dk}{a\left(  t\right)  }\rightarrow\mathcal{M}H\int dt
\end{equation}
instead of $\int dk\rightarrow eE_{0}\int dt$ like in the Minkowski space-time.

From equation (\ref{67}), we find that, for a pure gravitational field (i.e.
$eE_{0}=0$),%

\begin{equation}
\frac{dN}{dxdt}=\frac{mH}{\pi}\exp\left(  -\frac{2\pi}{H}m\right)  .
\label{69}%
\end{equation}
Furthermore, the number of created particles per unit of time per unit of
length can be written in the form%
\begin{equation}
\frac{dN}{dxdt}=\gamma_{E}\frac{mH}{\pi}\exp\left(  -\frac{2\pi}{H}m\right)  ,
\end{equation}
where the factor $\gamma_{E}$ is given by
\begin{equation}
\gamma_{E}=\sqrt{1+y^{2}}\cosh\left(  2\pi\frac{m}{H}y\right)  \exp\left[
\frac{2\pi m}{H}\left(  1-\sqrt{1+y^{2}}\right)  \right]  \label{gammae}%
\end{equation}
with%
\begin{equation}
y=\frac{eE_{0}}{mH}.
\end{equation}
It is easy to show that $\gamma_{E}$ is always greater that $1$ and therefore
the electric field amplifies the particle creation in de-Sitter space-time.

\begin{figure}
[ptb]
\begin{center}
\includegraphics[
height=4.7245in,
width=4.7245in
]%
{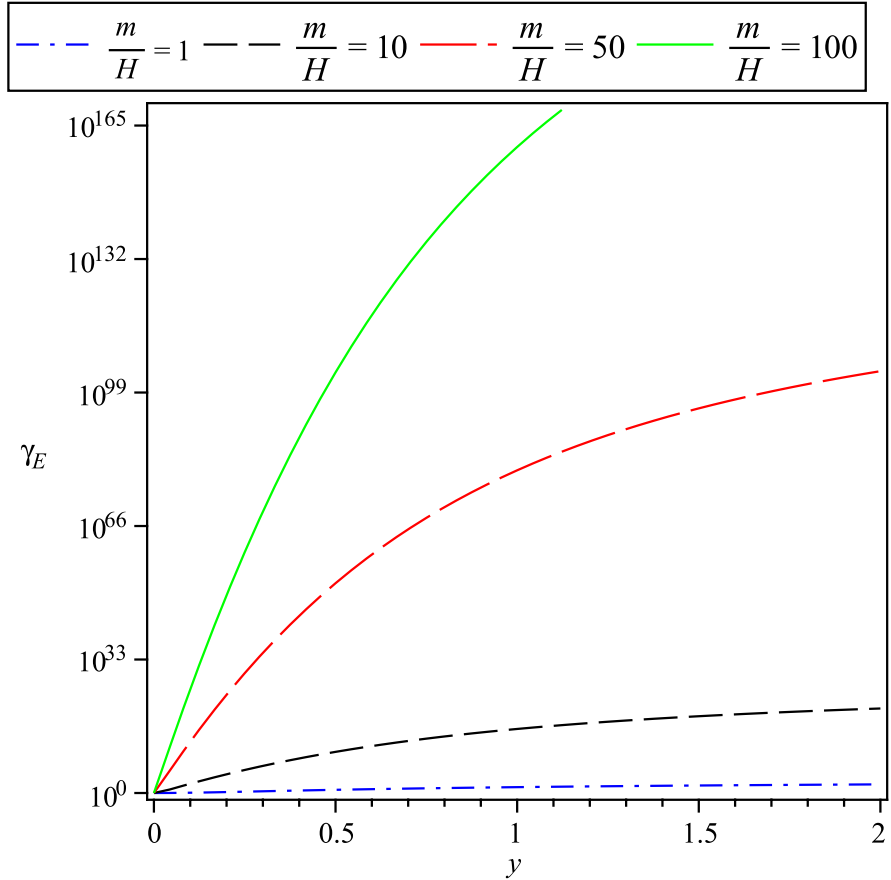}%
\caption{{\protect\small Plotting }$\gamma_{E}$ {\protect\small as a function
of the variable }$y=\frac{eE_{0}}{mH}${\protect\small  \ for various values of
}$\frac{m}{H}${\protect\small .}}%
\label{Fig1}%
\end{center}
\end{figure}

In figure (\ref{Fig1}), we plot the factor $\gamma_{E}$ as a function of the
variable $y=\frac{eE_{0}}{mH}$ for various values of $\frac{m}{H}$. As a
result, we remark that the electric field leads to a significant enhancement
of the particle creation. This effect is more important as soon as the
quantity $\frac{m}{H}$ is large.%

As is mentioned above, in a pure gravitational field, the production of
massive (heavy) particles is negligible because of the exponential
$\exp\left(  -\frac{2\pi}{H}m\right)  $ and the creation of massive particles
is excluded to play any role in the evolution of the early universe. However,
if electric fields existed in the early stages of the universe they could
enhance the creation of massive particles and, thus, the particle creation
would have significant implications on the cosmic evolution.

\section{Schwinger effective Lagrangian}

More recently the authors of \cite{2d} have computed the one-loop effective
Lagrangian for scalar particles in $dS_{2}$ space-time, starting from the
definition%
\begin{equation}
\mathcal{L}_{dS}^{(1)}=iH\int\frac{dk}{2\pi}\ln\left(  \alpha_{k}^{\ast
}\right)  ,
\end{equation}
where $\alpha_{k}^{\ast}$ is the Bogoliubov coefficient. They have found that
the imaginary part of this effective Lagrangian is given by
\begin{equation}
2\operatorname{Im}\mathcal{L}_{dS}^{(1)}=\frac{qE}{2\pi}\ln\left(
1+N_{ds}\right)  ,
\end{equation}
where $N_{ds}$ is the number density of created scalar particles. In that
derivation the integration over $k$ is replaced by the factor $eE$, like in
the Minkowski space, and the fact that the coefficient $\alpha_{k}^{\ast}$
depends on the sign of $k$ is not taken into account.

In this section we show how to derive the imaginary part of the Schwinger
effective Lagrangian from our previous results. We start by writing
$\mathcal{P}_{k}$ as
\begin{equation}
\mathcal{P}_{k}=\frac{\sigma}{1-\sigma}, \label{58}%
\end{equation}
where%
\[
\sigma=\frac{\sinh\pi\left(  \frac{\mathcal{M}}{H}+\epsilon\frac{eE_{0}}%
{H^{2}}\right)  e^{-2\pi\frac{\mathcal{M}}{H}}}{\sinh\pi\left(  \frac
{\mathcal{M}}{H}+\epsilon\frac{eE_{0}}{H^{2}}\right)  e^{-2\pi\frac
{\mathcal{M}}{H}}+\sinh\pi\left(  \frac{\mathcal{M}}{H}-\epsilon\frac{eE_{0}%
}{H^{2}}\right)  },
\]
where $\epsilon=\operatorname{sign}\left(  k\right)  $.

Let $\mathcal{C}_{k}$ to be the probability to have no pair creation in the
state $k.$ The quantity $\mathcal{C}_{k}\mathcal{P}_{k}$ is then the
probability to have only one pair in the state $k.$ Because of the Pauli
principle we have $\mathcal{C}_{k}+\mathcal{C}_{k}\mathcal{P}_{k}=1$ and%
\begin{equation}
\mathcal{C}_{k}=1-\sigma. \label{60}%
\end{equation}
Next, we define the vacuum to vacuum transition amplitude by an intermediate
effective action $\mathcal{A}\left(  vac-vac\right)  =\exp\left(
iS_{eff}\right)  $. The vacuum to vacuum probability can be then written as
\begin{equation}
\left\vert \mathcal{A}\left(  vac-vac\right)  \right\vert ^{2}=\exp\left(
-2\operatorname{Im}S_{eff}\right)  =\prod_{k}\mathcal{C}_{k}.
\end{equation}
It follows from (\ref{60}) that%

\begin{equation}
\exp\left(  -2\operatorname{Im}S_{eff}\right)  =\exp\left[  \sum_{k}\ln\left(
1-\sigma\right)  \right]  , \label{61}%
\end{equation}
where the sum $\sum_{k}$ has to be understood as $\int\frac{dxdk}{2\pi
a\left(  t\right)  }$. Taking into account that
\begin{equation}
\ln(1-\sigma)=\ln\left[  1-e^{-2\pi\left(  \frac{\mathcal{M}}{H}%
-\operatorname{sign}\left(  k\right)  \frac{eE_{0}}{H^{2}}\right)  }\right]
-\ln\left[  1-e^{-4\pi\frac{\mathcal{M}}{H}}\right]
\end{equation}
we obtain%
\begin{equation}
2\operatorname{Im}S_{eff}=-\int\frac{dxdk}{2\pi a\left(  t\right)  }\ln\left(
1-e^{-2\pi\left(  \frac{\mathcal{M}}{H}-\operatorname{sign}\left(  k\right)
\frac{eE_{0}}{H^{2}}\right)  }\right)  +\int\frac{dxdk}{2\pi a\left(
t\right)  }\ln\left(  1-e^{-4\pi\frac{\mathcal{M}}{H}}\right)  . \label{62}%
\end{equation}
By expanding the logarithm functions, we obtain%
\begin{align}
2\operatorname{Im}S_{eff}\  &  =2\int_{k\geq0}\frac{dxdk}{2\pi a\left(
t\right)  }\sum_{n=1}\frac{1}{n}\cosh\left(  2\pi n\frac{eE_{0}}{H^{2}%
}\right)  \exp\left(  -\frac{2\pi n}{H}\mathcal{M}\right) \nonumber\\
&  -2\int_{k\geq0}\frac{dxdk}{2\pi a\left(  t\right)  }\sum_{n=1}\frac{1}%
{n}\exp\left(  -\frac{4\pi n}{H}\mathcal{M}\right)
\end{align}
To eliminate the divergence arises in the integration over $k$, we define an
effective Lagrangian%
\[
2\operatorname{Im}S_{eff}=\int dxdt\ 2\operatorname{Im}L_{eff}%
\]
and we proceed like in the previous section. The\ quantity $2\operatorname{Im}%
L_{eff},$ which has to be interpreted as the probability of particle creation
per unit of time per unit of length, will be then given by
\begin{align}
2\operatorname{Im}L_{eff}  &  =\frac{\mathcal{M}H}{\pi}\sum_{n=1}\frac{1}%
{n}\exp\left(  -\frac{2\pi n}{H}\mathcal{M}\right)  \cosh\left(  2\pi
n\frac{eE_{0}}{H^{2}}\right) \nonumber\\
&  -\frac{\mathcal{M}H}{\pi}\sum_{n=1}\frac{1}{n}\exp\left(  -\frac{4\pi n}%
{H}\mathcal{M}\right)  .
\end{align}
Taking into account that%
\begin{equation}
\int_{-\infty}^{+\infty}\frac{ds}{s}\exp\left(  -i\frac{\delta}{2}s\right)
\left(  \coth\frac{s}{2}-\frac{2}{s}\right)  =\sum_{n=1}\frac{1}{n}\exp\left(
-\pi n\delta\right)
\end{equation}
we can write the imaginary part of the effective Lagrangian%
\begin{align}
2\operatorname{Im}L_{eff}  &  =\frac{\mathcal{M}H}{\pi}\int_{-\infty}%
^{+\infty}\frac{ds}{s}\exp\left(  -i\frac{\mathcal{M}}{H}s\right)  \cos\left(
\frac{eE_{0}}{H^{2}}s\right)  \left(  \coth\frac{s}{2}-\frac{2}{s}\right)
\nonumber\\
&  -\frac{\mathcal{M}H}{\pi}\int_{-\infty}^{+\infty}\frac{ds}{s}\exp\left(
-2i\frac{\mathcal{M}}{H}s\right)  \left(  \coth\frac{s}{2}-\frac{2}{s}\right)
\end{align}
Then the Schwinger-like effective Lagrangian is of the form%
\begin{align}
L_{eff}  &  =i\frac{\mathcal{M}H}{\pi}\int_{0}^{+\infty}\frac{ds}{s}%
\exp\left(  -i\frac{\mathcal{M}}{H}s\right)  \cos\left(  \frac{eE_{0}}{H^{2}%
}s\right)  \left(  \coth\frac{s}{2}-\frac{2}{s}\right) \nonumber\\
&  -i\frac{\mathcal{M}H}{\pi}\int_{0}^{+\infty}\frac{ds}{s}\exp\left(
-2i\frac{\mathcal{M}}{H}s\right)  \left(  \coth\frac{s}{2}-\frac{2}{s}\right)
+L_{0},
\end{align}
where $L_{0}$ is an unimportant real phase.

Notice that in the limit $E_{0}\rightarrow0$, we find%
\begin{equation}
L_{eff}=i\frac{mH}{\pi}\int_{0}^{+\infty}\frac{ds}{s}\exp\left(  -i\frac{m}%
{H}s\right)  \left(  1-\exp\left(  -i\frac{m}{H}s\right)  \right)  \left(
\coth\frac{s}{2}-\frac{2}{s}\right)  +L_{0},
\end{equation}
and%
\begin{equation}
2\operatorname{Im}L_{eff}=\frac{mH}{\pi}\sum_{n=1}\frac{1}{n}\exp\left(
-\frac{2\pi n}{H}m\right)  \left[  1-\exp\left(  -\frac{2\pi n}{H}m\right)
\right]  ,\nonumber
\end{equation}
which could be written in the form%
\begin{equation}
2\operatorname{Im}L_{eff}=\frac{mH}{\pi}\ln\left[  1+\exp\left(  -\frac{2\pi
}{H}m\right)  \right]  . \label{68}%
\end{equation}

\section{The weak expansion limit}

Let us now consider the weak expansion case, which is appropriate to the
actual accelerating phase of our universe. We concentrate our attention on how
the universe expansion influences on the Schwinger effect and how the usual
results corresponding to the flat space are reobtained. We discuss also the
possibility to enhance the Schwinger effect by the expansion of the universe.

\subsection{The number density of created particles}

Let us, first, study the number density of created particles when $H<<1$. In
order to obtain a constant electric field in the limit $H\rightarrow0$, where
the de-Sitter space reduces to the flat Minkowski space-time, we make the
shift
\[
A_{1}\rightarrow A_{1}^{\prime}=\frac{E_{0}}{H}e^{Ht}-\frac{E_{0}}{H}%
\]
In such a case, we have to replace $k$ by $k^{\prime}=k+\frac{eE_{0}}{H}$ in
order to obtain the pair production probability and the density of particles.
We get
\begin{equation}
n\left(  k\right)  =\left\{
\begin{array}
[c]{ccc}%
\exp\left[  -\frac{2\pi}{H}\left(  \mathcal{M}-\frac{eE_{0}}{H}\right)
\right]  & \text{ \ \ \ \ \ \ \ if \ \ \ \ \ \ \ } & k>-\frac{eE_{0}}{H}\\
\exp\left[  -\frac{2\pi}{H}\left(  \mathcal{M}+\frac{eE_{0}}{H}\right)
\right]  & \text{ \ \ \ \ \ \ \ \ if \ \ \ \ \ \ \ \ } & k<-\frac{eE_{0}}{H}%
\end{array}
\right.  \label{s}%
\end{equation}
It is then clear that the exponential in the second line of equation (\ref{s})
goes to $0$ as $H\rightarrow0$. For the first line, we use the limit
\[
\frac{\mathcal{M}}{H}-\frac{eE_{0}}{H^{2}}\simeq\frac{m^{2}}{2eE_{0}}-\frac
{1}{8}\frac{m^{4}}{e^{3}E_{0}^{3}}H^{2}%
\]
to obtain%
\begin{equation}
\mathcal{P}_{k}\approx\frac{\exp\left[  -\pi\left(  \frac{\mathcal{M}}%
{H}-\frac{eE_{0}}{H^{2}}\right)  \right]  }{2\sinh\pi\left(  \frac
{\mathcal{M}}{H}-\frac{eE_{0}}{H^{2}}\right)  }\approx\frac{\exp\left[
-\pi\left(  \frac{m^{2}}{eE_{0}}-\frac{1}{4}\frac{m^{4}}{e^{3}E_{0}^{3}}%
H^{2}\right)  \right]  }{1-\exp\left[  -\pi\left(  \frac{m^{2}}{eE_{0}}%
-\frac{1}{4}\frac{m^{4}}{e^{3}E_{0}^{3}}H^{2}\right)  \right]  }%
\end{equation}
and
\begin{equation}
n\left(  k\right)  \approx\exp\left[  -\pi\left(  \frac{m^{2}}{eE_{0}}%
-\frac{1}{4}\frac{m^{4}}{e^{3}E_{0}^{3}}H^{2}\right)  \right]  =\exp\left(
\frac{\pi}{4}\frac{m^{4}}{e^{3}E_{0}^{3}}H^{2}\right)  \exp\left(  -\pi
\frac{m^{2}}{eE_{0}}\right)  .
\end{equation}
Besides, when $H\rightarrow0$, we see that%
\begin{equation}
n\left(  k\right)  \approx\exp\left(  -\pi\frac{m^{2}}{eE_{0}}\right)
\label{56}%
\end{equation}
and
\begin{equation}
\mathcal{P}_{k}\approx\frac{\exp\left(  -\pi\frac{m^{2}}{eE_{0}}\right)
}{1-\exp\left(  -\pi\frac{m^{2}}{eE_{0}}\right)  }. \label{57}%
\end{equation}
Let us notice that the condition $k>-\frac{eE_{0}}{H}$ means that, in the
limit $H\rightarrow0$, we have the possibility to create particles with an
arbitrary wave vector, $k\in\left]  -\infty,+\infty\right[  $. Physically,
this fact is not strange because $k$ is the canonical momentum and not the
physical one. We remark also that equations (\ref{56}) and (\ref{57}) are in
complete agreements with the well-known results corresponding to the pair
creation by an electric field in Minkowski space-time (see \cite{CF1} and
references therein).

\subsection{The total number of created particles}

In Minkowski space time, the number of created particles per unit of time per
unit of length is given by%
\begin{equation}
\frac{dN}{dxdt}=\frac{eE_{0}}{2\pi}\exp\left(  -\pi\frac{m^{2}}{eE_{0}%
}\right)  .
\end{equation}
Because of the exponential $\exp\left(  -\pi\frac{m^{2}}{eE_{0}}\right)  $,
the field strength required to see the Schwinger effect is of order of the
critical value $E_{c}=10^{16}\mathtt{Vcm}^{-1}$. However, the maximal electric
field produced by the current technologies is about two orders of magnitude
smaller than the critical value. This makes$\frac{dN}{dxdt}$ very small for a
pure electric field. Then, in order to see the effect the $\frac{dN}{dxdt}$
must be enhanced exponentially by compensating $\exp\left(  -\pi\frac{m^{2}%
}{eE_{0}}\right)  $.

We consider an electric field $E_{0}$ of the order $\sim10^{-2}E_{c}$.
Therefore, the present day expansion satisfies the inequality%
\begin{equation}
\frac{eE_{0}}{H^{2}}>>\frac{m^{2}}{eE_{0}}. \label{c}%
\end{equation}
In this limit we have%

\begin{equation}
\frac{dN}{dxdt}=\frac{\mathcal{M}H}{\pi}\exp\left[  -\frac{2\pi}{H}\left(
\mathcal{M}-\frac{eE_{0}}{H}\right)  \right]  ,
\end{equation}
which can be approximated by%
\begin{equation}
\frac{dN}{dxdt}=\gamma_{H}\frac{eE_{0}}{2\pi}\exp\left(  -\pi\frac{m^{2}%
}{eE_{0}}\right)  ,
\end{equation}
where%
\begin{equation}
\gamma_{H}=\sqrt{1+\frac{m^{2}H^{2}}{e^{2}E_{0}^{2}}}\exp\left(  \pi
\frac{m^{4}H^{2}}{4e^{3}E_{0}^{3}}\right)  .
\end{equation}
From the condition (\ref{c}), we find that the $\frac{m^{4}H^{2}}{4e^{3}%
E_{0}^{3}}<<\frac{m^{2}}{eE_{0}}$ and then the Schwinger effect cannot be
exponentially enhanced by the expansion of the universe.

In recent years, some explicit experimental realizations are proposed to
observe the Schwinger effect for the first time \cite{exp1,exp2,exp3}. The
basic principle of these experiences is the dynamic assistance of the
Schwinger mechanism by the combination of slower pulse strong Laser with a
faster pulse weak one. It this case, the faster pulse gives a multi-photon
contribution which leads to an exponential enhancement. Since $\gamma
_{H}\simeq1$, even if the Schwinger effect is observed in laboratory, it would
be difficult to deduce from this observation whether we live in an expanding
universe or in a Minkowski space-time.

\subsection{The Schwinger effective Lagrangian}

In weak expansion limit the\ probability of particle creation per unit of time
per unit of length will be then given by
\begin{equation}
2\operatorname{Im}L_{eff}=\frac{\mathcal{M}H}{2\pi}\sum_{n=1}\frac{1}{n}%
\exp\left[  -\frac{2\pi n}{H}\left(  \mathcal{M}-\frac{eE_{0}}{H}\right)
\right]  \label{we}%
\end{equation}
which can be written as an integral over the Schwinger proper time
\begin{equation}
2\operatorname{Im}L_{eff}=\frac{\mathcal{M}H}{2\pi}\int_{-\infty}^{+\infty
}\frac{ds}{s}\exp\left[  -i\left(  \frac{\mathcal{M}}{H}-\frac{eE_{0}}{H^{2}%
}\right)  s\right]  \left(  \coth\frac{s}{2}-\frac{2}{s}\right)
\end{equation}
Then the Schwinger-like effective Lagrangian is of the form
\begin{equation}
L_{eff}=\frac{i}{2\pi}\frac{\mathcal{M}H}{eE_{0}}\int_{0}^{+\infty}\frac
{ds}{s}\exp\left[  -i\left(  \frac{\mathcal{M}}{H}-\frac{eE_{0}}{H^{2}%
}\right)  s\right]  \left(  \coth\frac{s}{2}-\frac{2}{s}\right)  +...
\end{equation}
By taking the limit $H\rightarrow0$ and making the change $s\rightarrow
2eE_{0}s$, we obtain the well-known result
\begin{equation}
L_{eff}=\frac{i}{2\pi}\int_{0}^{+\infty}\frac{ds}{s}\exp\left(  -im^{2}%
s\right)  \left(  eE_{0}\coth\left(  eE_{0}s\right)  -\frac{1}{s}\right)  +...
\end{equation}
Doing integration over $s$ by using the residues theorem or taking the limit
$H\rightarrow0$ in equation (\ref{we}), we obtain
\begin{equation}
2\operatorname{Im}L_{eff}=\frac{eE_{0}}{2\pi}\sum_{n=1}\frac{1}{n}\exp\left(
-n\pi\frac{m^{2}}{eE_{0}}\right)  ,
\end{equation}
which is the same as equation (64) in \cite{Lin}.

\section{Concluding remarks}

In conclusion, we have studied in this paper the influence of an electric
field on the creation of spin-$\frac{1}{2}$ particle pairs in the (1+1)
inflationary de-Sitter space-time. We have shown at the first stage that the
Schwinger effect in an expanding universe is effectively equivalent to the
Schwinger effect in Minkowski space-time by introducing an unitary
transformation to the Dirac equation. This transformation allows us to obtain
exact solutions for the case of de-Sitter space with a constant electric
field. The charge symmetry between positive and negative frequency modes
permitted us to find exact and analytic expressions for the Bogoliubov
coefficients without need to normalize the wave functions. Then the
probability to create a pair of particles in a given state and the density of
created fermions are calculated. The obtained expressions show that a constant
electric field increases the creation of fermions with positive wave number
$k$ and minimizes it in the opposite direction. This effect, which depends on
the particle charge and the orientation of the electric field, is of interest
because in the absence of the electric field there is no preferred direction
for the created particles. In addition, we have shown that the creation of
massless particles with conformal coupling is impossible even if an electric
field is present.

By doing summation over all allowed states we have expressed the number of
created particles per unit of time per unit of length and the imaginary part
of the Schwinger effective Lagrangian in closed forms. We have shown that the
electric field leads to a significant enhancement of the particle creation.
Then, if the particle creation effects on the cosmic evolution are negligible
in a pure gravitational field, the presence of a strong electric field makes
these effects appreciable. Therefore, the electromagnetic fields influence on
the cosmic evolution directly via Friedmann equations and by their effect on
the creation of particles.

We have also considered the case of weak expansion and we have discussed the
effect of the actual expansion of the universe on the Schwinger effect. It is
shown that the universe expansion could not assist the Schwinger effect to be
observed in laboratory and even if the effect is enhanced by the combination
of stronger slower pulse with a faster weaker pulse, it would be difficult to
see through this effect if the universe is expanding.

Even if the problem is modeled in a two dimensional space-time it shows
clearly the effect of the electric field on the creation of fermions. Taking
into account that the number of created particles per unit of time and length
in a pure gravitational field, is given, in the case of (3 + 1) dimensional
space-time, by \cite{ds4}%
\[
\frac{dN}{dxdt}=\frac{m^{3}H}{\pi^{2}}~\exp\left(  -\frac{2\pi m}{H}\right)
,
\]
we expect that the four dimensional analogue of equation (\ref{67}) is of the form%

\begin{equation}
\frac{dN}{dxdt}=\frac{\mathcal{M}^{3}H}{\pi^{2}}~F\left(  \frac{eE_{0}}{H^{2}%
}\right)  \exp\left(  -\frac{2\pi}{H}\mathcal{M}\right)  ,
\end{equation}
where the function $F\left(  \frac{eE_{0}}{H^{2}}\right)  $ satisfies the
condition $F\left(  0\right)  =1$. If we impose that this expression reduces
to the Schwinger result \cite{CF1}%
\begin{equation}
\frac{dN}{dxdt}=\frac{e^{2}E_{0}^{2}}{2\pi^{2}}~\exp\left(  -\pi\frac{m^{2}%
}{eE_{0}}\right)  ,
\end{equation}
when $H\rightarrow0$, the function $F\left(  \frac{eE_{0}}{H^{2}}\right)  $
must behaves, for $\frac{eE_{0}}{H^{2}}>>1$, like%
\begin{equation}
F\left(  \frac{eE_{0}}{H^{2}}\right)  \simeq\frac{H^{2}}{4\pi eE_{0}}%
\exp\left(  2\pi\frac{eE_{0}}{H^{2}}\right)  .
\end{equation}
This implies that the amplification factor in (3+1) dimensions is
exponentially equivalent to $\gamma_{E}$ given in (\ref{gammae}).

\appendix

\section{The relativistic Hamilton-Jacobi equation}

The "\textit{in}" and "\textit{out}" states are chosen according to the
asymptotic behavior of the semiclassical solutions constructed by the use of
WKB approximation starting from the solutions of the relativistic
Hamilton-Jacobi (H-J) equation. In this section, we give the analytic solution
of the H-J equation. The general form of the H-J equation is given by%

\begin{equation}
g^{\mu\nu}\left(  \partial_{\mu}S+eA_{\mu}\right)  \left(  \partial_{\nu
}S+eA_{\nu}\right)  -m^{2}=0 \label{a1}%
\end{equation}
For the case of $dS_{2}$ space with constant electric field the classical
action $S$ can be decomposed as%
\begin{equation}
S=G\left(  \eta\right)  +kx, \label{a2}%
\end{equation}
where the time dependent part $G\left(  \eta\right)  $ satisfies the following equation%

\begin{equation}
\frac{\partial}{\partial\eta}G\left(  \eta\right)  =\pm\sqrt{\left(
k+\frac{eE_{0}}{H^{2}\eta}\right)  ^{2}+\frac{m^{2}}{H^{2}\eta^{2}}}.
\label{a3}%
\end{equation}
By the use of the well-known integral \cite{Grad}
\begin{align}
&  \left.  \int\frac{\sqrt{\alpha+\beta x+\gamma x^{2}}}{x}dx=\right.
\nonumber\\
&  \sqrt{\alpha+\beta x+\gamma x^{2}}-\sqrt{\alpha}\operatorname{arcsinh}%
\left(  \frac{2\alpha+\beta x}{x\sqrt{4\alpha\gamma-\beta^{2}}}\right)
\nonumber\\
&  +\frac{\beta}{2}\frac{1}{\sqrt{\gamma}}\operatorname{arcsinh}\left(
\frac{2\gamma x+\beta}{\sqrt{4\alpha\gamma-\beta^{2}}}\right)  , \label{a4}%
\end{align}
where $\alpha$, $\beta$ and $\gamma$ are real numbers, with $\gamma>0$ and
$4\alpha\gamma-\beta^{2}>0$, we obtain%

\begin{align}
&  \left.  G\left(  \eta\right)  =\pm\sqrt{k^{2}\eta^{2}+2k\frac{eE_{0}}%
{H^{2}}\eta+\frac{\mathcal{M}^{2}}{H^{2}}}\right. \nonumber\\
&  \mp\frac{\mathcal{M}}{H}\ln\left(  \frac{\frac{\mathcal{M}^{2}}{H^{2}%
}+k\frac{eE_{0}}{H^{2}}\eta}{\frac{m}{H}k\eta}+\sqrt{\left(  \frac
{\frac{\mathcal{M}^{2}}{H^{2}}+k\frac{eE_{0}}{H^{2}}\eta}{\frac{m}{H}k\eta
}\right)  ^{2}+1}\right) \nonumber\\
&  \pm\frac{eE_{0}}{H^{2}}\ln\left(  k\eta+\frac{eE_{0}}{H^{2}}+\sqrt{\left(
k\eta+\frac{eE_{0}}{H^{2}}\right)  ^{2}+\left(  \frac{m}{H}\right)  ^{2}%
}\right)  . \label{a5}%
\end{align}
Notice that when $k\eta<<\frac{\mathcal{M}}{H}$, we have $G\left(
\eta\right)  \approx\pm\mathcal{M}t+constant$.

\section{Useful relations}

The full spinor structure is obtained by the use of the following relations
\cite{Vilenkin}%
\begin{equation}
\left[  \left(  2\mu-1\right)  \frac{\partial}{\partial\rho}+\frac{\left(
2\mu-1\right)  ^{2}}{2\rho}-\lambda\right]  W_{\lambda,\mu}\left(
\rho\right)  =-\left(  \mu+\lambda-\frac{1}{2}\right)  W_{\lambda,\mu
-1}\left(  \rho\right)  \label{29}%
\end{equation}
and%
\begin{equation}
\left[  \left(  2\mu+1\right)  \frac{\partial}{\partial\rho}-\frac{\left(
2\mu+1\right)  ^{2}}{2\rho}+\lambda\right]  M_{\lambda,\mu}\left(
\rho\right)  =\frac{\left(  \mu+\frac{1}{2}\right)  ^{2}-\lambda^{2}}{2\left(
\mu+1\right)  \left(  2\mu+1\right)  }M_{\lambda,\mu+1}\left(  \rho\right)  .
\label{39}%
\end{equation}

Taking into account that $W_{\lambda,-\mu}\left(  \rho\right)  =W_{\lambda
,\mu}\left(  \rho\right)  $ and $\left(  -\rho\right)  ^{-\frac{1}{2}-\mu
}M_{-\lambda,\mu}\left(  -\rho\right)  =\left(  \rho\right)  ^{-\frac{1}%
{2}-\mu}M_{\lambda,\mu}\left(  \rho\right)  $ and using the following
functional relations \cite{Vilenkin}%

\begin{equation}
W_{\lambda,\mu}\left(  \rho\right)  +\frac{1}{\sqrt{\rho}}\left[  \left(
\mu-\lambda+\frac{1}{2}\right)  W_{\lambda-\frac{1}{2},\mu+\frac{1}{2}}\left(
\rho\right)  -W_{\lambda+\frac{1}{2},\mu+\frac{1}{2}}\left(  \rho\right)
\right]  =0,
\end{equation}%
\begin{equation}
W_{\lambda,\mu}\left(  \rho\right)  -\frac{1}{\sqrt{\rho}}\left[  \left(
\mu+\lambda-\frac{1}{2}\right)  W_{\lambda-\frac{1}{2},\mu-\frac{1}{2}}\left(
\rho\right)  +W_{\lambda+\frac{1}{2},\mu-\frac{1}{2}}\left(  \rho\right)
\right]  =0,
\end{equation}
$\allowbreak$%
\begin{equation}
M_{\lambda,\mu}\left(  \rho\right)  +\frac{2\mu}{\sqrt{\rho}}\left(
M_{\lambda+\frac{1}{2},\mu-\frac{1}{2}}\left(  \rho\right)  -M_{\lambda
-\frac{1}{2},\mu-\frac{1}{2}}\left(  \rho\right)  \right)  =0,
\end{equation}
and%
\begin{equation}
M_{\lambda,\mu}\left(  \rho\right)  =\frac{1}{\left(  2\mu+1\right)
\sqrt{\rho}}\left[  \left(  \mu-\lambda+\frac{1}{2}\right)  M_{\lambda
-\frac{1}{2},\mu+\frac{1}{2}}\left(  \rho\right)  +\left(  \mu+\lambda
+\frac{1}{2}\right)  M_{\lambda+\frac{1}{2},\mu+\frac{1}{2}}\left(
\rho\right)  \right]
\end{equation}
we make connection with the solutions reported in \cite{2f} and \cite{2g}.

\end{document}